\documentclass[a4paper,11pt]{article}
\usepackage{pos}

\title{Theoretical interpretation of the observed neutrino emission from Tidal Disruption Events}
 \ShortTitle{TDE neutrino interpretation}

\author*[a]{Walter Winter}
\author*[b]{Cecilia Lunardini}

\affiliation[a]{
Deutsches Elektronen-Synchrotron (DESY),\\
Platanenallee 6, D-15738,
Zeuthen, Germany}

\affiliation[b]{
Arizona State University, Department of Physics, \\
450 E. Tyler Mall, Tempe, AZ 85287-1504 USA}

\emailAdd{walter.winter@desy.de}
\emailAdd{Cecilia.Lunardini@asu.edu}

\abstract{The observation of a neutrino at IceCube in association with the Tidal Disruption Event (TDE) AT2019dsg, has suggested TDEs as a new class of sources of astrophysical neutrinos. We present a model of this multi-messenger observation in a jetted concordance scenario, where the neutrino production is directly linked to the observed X-rays, and  the timing of the neutrino observation (about 150 days post peak) can be naturally described. We briefly discuss the implications of our model for future neutrino-TDE associations.}

\FullConference{37$^{\rm{th}}$ International Cosmic Ray Conference (ICRC 2021)\\
		July 12th -- 23rd, 2021\\
		Online -- Berlin, Germany}

\begin{document}
\maketitle

\section{Introduction}

\nocite{Winter:2020ptf}

In 2019, an IceCube neutrino event (IC191001A)  has been associated with the Tidal Disruption Event (TDE) AT2019dsg, with a probability of random coincidence of 0.2\% to 0.5\%~\cite{Stein:2020xhk}; the neutrino was detected at  $t \simeq 150$ days after the optical peak of the candidate parent TDE. AT2019dsg was found to be bright in X-rays, which were detected 17 days after peak time with a thermal spectrum with $T_X = 60 \, \mathrm{eV}$ and quickly decayed over a timescale of tens of days. In addition,  radio emission -- nearly  constant over 90 days -- was found, and was interpreted as due to a mildly relativistic outflow.
Theoretical interpretations speculate that the abundant UV photons could be a target for the neutrino production~\cite{Stein:2020xhk},
whereas other possible production sites include a disk corona or hidden winds, possibly induced by collisions among tidal streams~\cite{Murase:2020lnu}, or the disk itself~\cite{Hayasaki:2019kjy}. 
Our recent work, Ref.~\cite{Winter:2020ptf}, gives an interpretation in terms of a relativistic jet, which could be be off-axis (see~\cite{Liu:2020isi}); we refer to~\cite{Hayasaki:2021jem} for an overview of all models.

In this proceeding we present our jetted model, Following Ref.~\cite{Winter:2020ptf}.  To lay the context,  let us overview the 
basic ingredients of the neutrino production 
which 
that can be inferred from data. First, the observed (likely) neutrino energy, $E_\nu \simeq 200$~TeV, implies a primary cosmic-ray  energy of at least 4~PeV, which means that the production site must be an efficient cosmic-ray accelerator. 
Second, the 150 days delay of the neutrino detection requires 
that the neutrino fluence be sustained over a similarly long period of time. 
Both these aspects find a natural interpretation in our model (whereas they may be difficult to justify in other scenarios; for example acceleration might be insufficient in corona models \cite{Inoue:2021tcn}, and the delay  may be difficult to justify in models where the UV photon are the targets for neutrino production).
Specifically,  we propose a mechanism where photohadronic neutrino production takes place in internal shocks in a jet; there, the most likely  target photon energy to produce a 200~TeV neutrino is around 100~eV, close to the observed X-ray temperature of AT2019dsg.\footnote{See Fig. 4 in \cite{Hummer:2010vx}; estimate holds for externally production radiation to a jet or a mildly relativistic outflow.} Therefore, we propose that X-rays be the target; this has the advantage of only very mild requirements on the acceleration efficiency of the primaries, as will be seen in the reminder of this proceeding. 
In our proposed  mechanism, the time delay is related to the size of the newly formed system. 
A further argument in favor of a relativistic jet is energetic (see next section),  whereas the main argument against is the lack of direct jet signatures (addressed in Sec.~\ref{sec:discussion}). 

\section{An upper limit for the neutrino production energetics}
\label{sec:energy}

There are two basic constraints on the energetics of the neutrino production. First, the mass of the disrupted star imposes an upper limit for the available energy, about $10^{54} \, \mathrm{erg}$ for a solar mass-like star. This energy will be processed through the supermassive black hole (SMBH) into different components (thermal radiation, outflow, wind, jet, etc). Second, the Eddington luminosity $L_\mathrm{Edd} \simeq 10^{44} \, M_{\mathrm{SMBH}}/(10^6 \, M_\odot)$ is an estimate for how much energy can be re-processed through accretion (although super-Eddington luminosities can be realized at peak). 

Assuming an average mass accretion rate of $\dot M=25 \, L_{\mathrm{Edd}}$ (sustained over hundreds of days), an upper limit for the neutrino luminosity is obtained:
\begin{equation}
L_\nu \sim 25 \, L_{\mathrm{Edd}} \, f_{\mathrm{comp}} \, \varepsilon_{\mathrm{acc}} \, \tau \, \frac{1}{8} \ll 0.1 \, L_{\mathrm{Edd}}~.
\label{equ:neutrino}
\end{equation}
Here $f_{\mathrm{comp}} \lesssim 0.2$ is the fraction of energy going into the component  where the neutrino is produced (in our model, the jet), $\varepsilon_{\mathrm{acc}} \ll 0.2$ is the fraction of energy converted into non-thermal protons at PeV energies\footnote{This contains actually two factors: the fraction converted into non-thermal protons, and a spectral index-dependent bolometric correction because only a small fraction of particles will reach the highest energies.}, $\tau \le 1$ is the (effective) optical thickness for the interactions, and $1/8$ ($1/2$ into charged pions, decaying into four leptons each) is the fraction of energy going into each neutrino flavor. These numerical values lead to the maximum value $ 0.1 L_{\mathrm{Edd}}$ in Eq. (\ref{equ:neutrino}), which roughly holds for both $pp$ and $p\gamma$ interactions.  Note that  the assumed values are rather optimistic here  (e.g.,  typically $\tau \ll 1$), which means that the actual neutrino luminosity is likely to be much smaller. From Eq. (\ref{equ:neutrino}), a bound on the total energy emitted in neutrinos follows: 
\begin{equation}
    E^{tot}_\nu \lesssim 200 \, \mathrm{days} \, 0.1 \, L_{\mathrm{Edd}} \simeq 2 \, 10^{50} \, \mathrm{erg} \,  \frac{M_{\mathrm{SMBH}}}{10^6 \, M_\odot} \, , \label{equ:enu}
\end{equation}
which is consistent with the estimate in \cite{Stein:2020xhk}, and corresponds to $N_\nu \simeq 0.2$ neutrino events at IceCube for $M_{\mathrm{SMBH}} = 10^6 \, M_\odot$ for the anticipated spectral shape~\cite{Fiorillo:2021hty}. As follows from Eq.~(\ref{equ:enu}), the neutrino production depends strongly on $M_{\mathrm{SMBH}}$. For AT2019dsg estimates of this quantity  are method-dependent,  varying between $1.3 \, 10^6 \, M_\odot$~\cite{Ryu:2020gxf} and $3 \, 10^7 \, M_\odot$~\cite{Stein:2020xhk} ; see also~\cite{Cannizzaro:2020xzc} for intermediate values. Considering this uncertainty, our arguments above leave two possible conclusions:
\begin{enumerate}
    \item 
     Either
     $M_{\mathrm{SMBH }} > 10^7 \, M_\odot$, \emph{and} the energy can be super-efficiently converted into neutrinos,
     \item
      or, the outflow where neutrino production occurs must be highly collimated, boosting $L_\nu \rightarrow L_\nu/\theta^2$, where $\theta$ is the opening angle; for a relativistic jet one may estimate that $\theta \simeq 1/\Gamma$. 
\end{enumerate}
Here we follow the second option, and use  $M_{\mathrm{SMBH }} =10^6 \, M_\odot$, which is energetically conservative (Eq. (\ref{equ:enu})). Many other models rely on alternative 1, typically with larger values of $M_{\mathrm{SMBH }}$.

\section{A jetted concordance scenario}

We note that the AT2019dsg observations -- such as the bolometric luminosity, $L_{\mathrm{bol}}$, the blackbody (BB) radius of optical-UV emission,  $R_{\mathrm{BB}}\simeq 5\, 10^{14}\mathrm{cm}$, and the X-ray emission --  are consistent with the TDE unified model in Ref.~\cite{Dai:2018jbr}, which is based on magneto-hydrodynamical (MHD) simulations. The model predicts a mildly relativistic outflow, consistent with AT2019dsg radio observations, and a relativistic jet for high enough SMBH spins, with a luminosity of about 20\% of the mass accretion rate. The X-rays are only visible in the ``funnel'' along the direction of angular momentum (and close to a possible jet). We assume 
that AT2019dsg 
is a realization of the unified model, 
observed in (or close to) the direction of the funnel;  we postulate the existence of a jet, which should be observable on- or slightly off-axis. In the region where plasma shells collide in the jet, internal shocks will accelerate protons to high energies. The typical collision radius is $R_C \simeq 2 \, \Gamma^2 \, t_v$. If the intermittent timescale of the engine $t_v$ is of the order of  the Schwarzschild time and $\Gamma \sim 7$, we get $R_C \sim R_{\mathrm{BB}}$.  We  therefore make the ansatz that $R_{C} \simeq R_{\mathrm{BB}}$ at all times, which implies that $R_{\mathrm{BB}}$  decreases slightly over time (following the observed trend for $R_{C}$); thus  enhancing the late-time neutrino production (since the production  efficiency scales like $ R_C^{-2}$). Considering that non-thermal radiation from the jet has not been established, we assume that is is sub-threshold (possible if, e.g., the jet baryonic loading  is high enough). See Ref.~\cite{Winter:2020ptf} for details of the model.

\begin{figure}[t]
    \centering
    \includegraphics[width=0.6\textwidth]{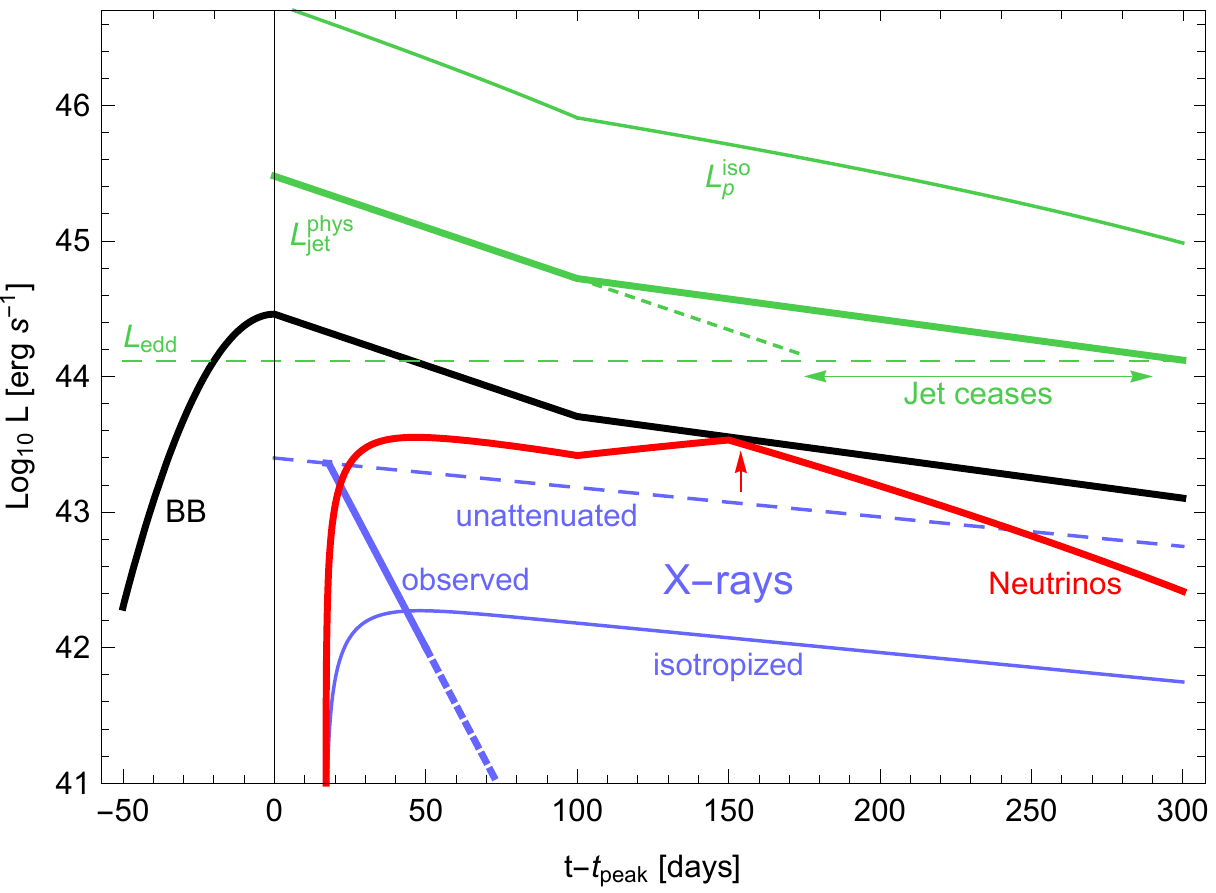}
    \caption{Evolution of luminosities as a function of time. The BB and X-ray luminosities (thick black and blue lines) have been observed, the neutrino luminosity (red curve) is our result, all other curves are assumptions of the model. The observed arrival time of the neutrino is marked by the arrow. The observed X-ray luminosity  decays quickly, whereas the unattenuated and isotropized contributions (shown for the the same energy range, relevant to the Swift observations) decline more slowly.  Figure adapted from Ref.~\cite{Winter:2020ptf}.  
    }
    \label{fig:lumicurves}
\end{figure}

The time evolution of the model ingredients and results are summarized in Fig.~\ref{fig:lumicurves}. The physical jet luminosity follows the BB luminosity and was normalized according to Ref.~\cite{Dai:2018jbr}; the jet ceases if it drops below $L_\mathrm{Edd}$ (at a time that is uncertain, but in any case after the observation of the neutrino, see arrow). The isotropic-equivalent proton luminosity is  $L_p^{\mathrm{iso}} \simeq (2 \Gamma^2) \varepsilon \, L_{\mathrm{jet}}^{\mathrm{phys}}$,  where $2 \Gamma^2$ is relativistic beaming factor (see Sec.~\ref{sec:energy}), and $\varepsilon \simeq 0.2$ is the fraction of energy that is transferred into non-thermal protons at all energies. Note that $\varepsilon_{\mathrm{acc}}  \ll  \varepsilon$ , cf.,  Eq.~(\ref{equ:neutrino}), because it only refers to PeV energy protons (i.e., contains the bolometric correction).

As indicated earlier, the thermal  X-rays from the accretion disk  serve as external target photons. A key observation of AT2019dsg is a quickly fading X-ray luminosity (solid blue curve in Fig.~\ref{fig:lumicurves}); exponentially decaying over the timescale of tens of days. The origin of the decay may be the outflow obscuring the X-rays --   consistent with 
Ref.~\cite{Dai:2018jbr} in terms of the optical thickness at the radii of interest -- or the cooling of the accretion disk, which shifts the peak of the X-ray spectrum out of the Swift energy window, see Ref.~\cite{Cannizzaro:2020xzc}, or both. We anticipate that a small fraction ($\sim $10\%) of the X-rays isotropize in the region relevant for the neutrino production, see Fig.~\ref{fig:lumicurves}, where the time evolution of the unattenuated X-rays follows the model in  \cite{Wen:2020cpm}. This fraction is somewhat {\em ad hoc}, as it includes both the covering factor and a possible dilution factor, depending on where these X-rays precisely isotropize.
Despite this uncertainty, however,  the isotropized flux can be considered realistic,  considering that the normalization of the unattenuated X-ray luminosity to the observed one 17 days after peak (Fig.~\ref{fig:lumicurves}) is very conservative. Indeed, the X-ray luminosity may have been higher at peak times (where no observations exist), perhaps as high as $L_{\mathrm{Edd}}$. Another interesting feature of the model is that the isotropization timescale is related to the observed X-ray decay timescale if it comes from obscuration, which is a again a measure for the propagation time of the outflow and for the size of the newly-built system; therefore, in our scenario no neutrinos are expected near the time of the optical peak. Note that a moderate cooling of the accretion disk would not affect our results significantly.
The resulting time evolution of the neutrino luminosity is given as red curve in Fig.~\ref{fig:lumicurves}: the average predicted arrival time is about 120  days post-peak, 30 days  earlier than the actual neutrino detection time.

\begin{figure}[t]
    \centering
    \includegraphics[width=0.4\textwidth]{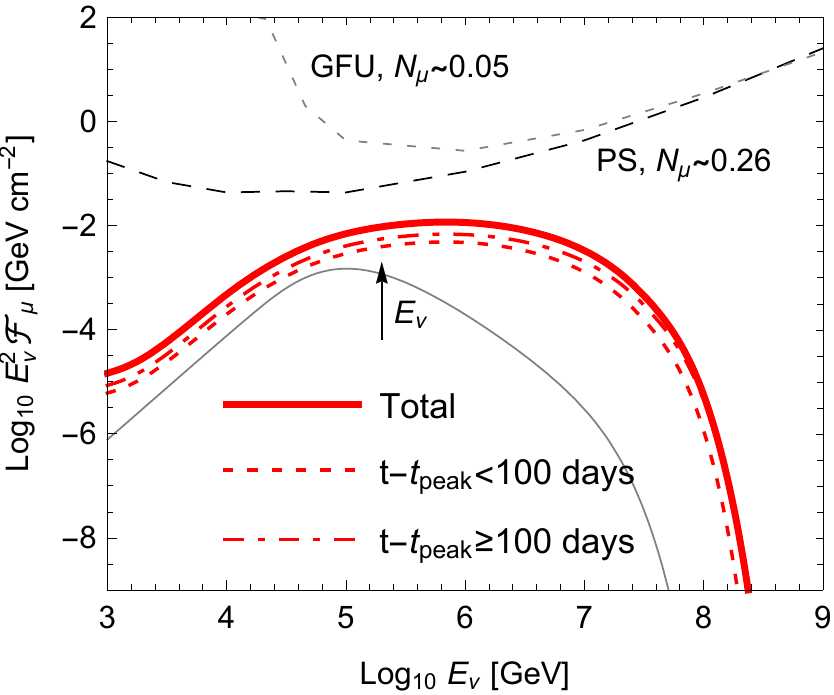}
    \caption{Predicted neutrino fluence as a function of energy, with contributions from early and late times (see legend), in comparison to the gamma-ray follow-up (GFU) and point source (PS) differential limits; the predicted event rates are given as well. We also show the $\Delta$-resonance contribution separately (gray curve). Figure adapted from Ref.~\cite{Winter:2020ptf}.  }
    \label{fig:spectrum}
\end{figure}

The predicted neutrino energy spectrum (multiplied by the square of the energy, $E_\nu^2 \mathcal{F}_\mu$) is given in Fig.~\ref{fig:spectrum}, with the likely (observed) neutrino energy indicated by the arrow. The curve is relatively flat over a large energy range, and the maximal energy is determined by the proton acceleration efficiency. The spectrum is wider compared to the frequently used $\Delta$-resonance approximation (gray curve), 
because multi-pion processes enhance the neutrino production at higher energies and lead to a neutrino spectral shape that follows more closely the primary proton spectrum~\cite{Fiorillo:2021hty}. The expected number of neutrinos events at IceCube is $N_\nu \simeq 0.05 - 0.26$ events, depending on the effective area used.

\section{Discussion}
\label{sec:discussion}

\begin{figure}[t]
    \centering
    \includegraphics[width=0.55\textwidth]{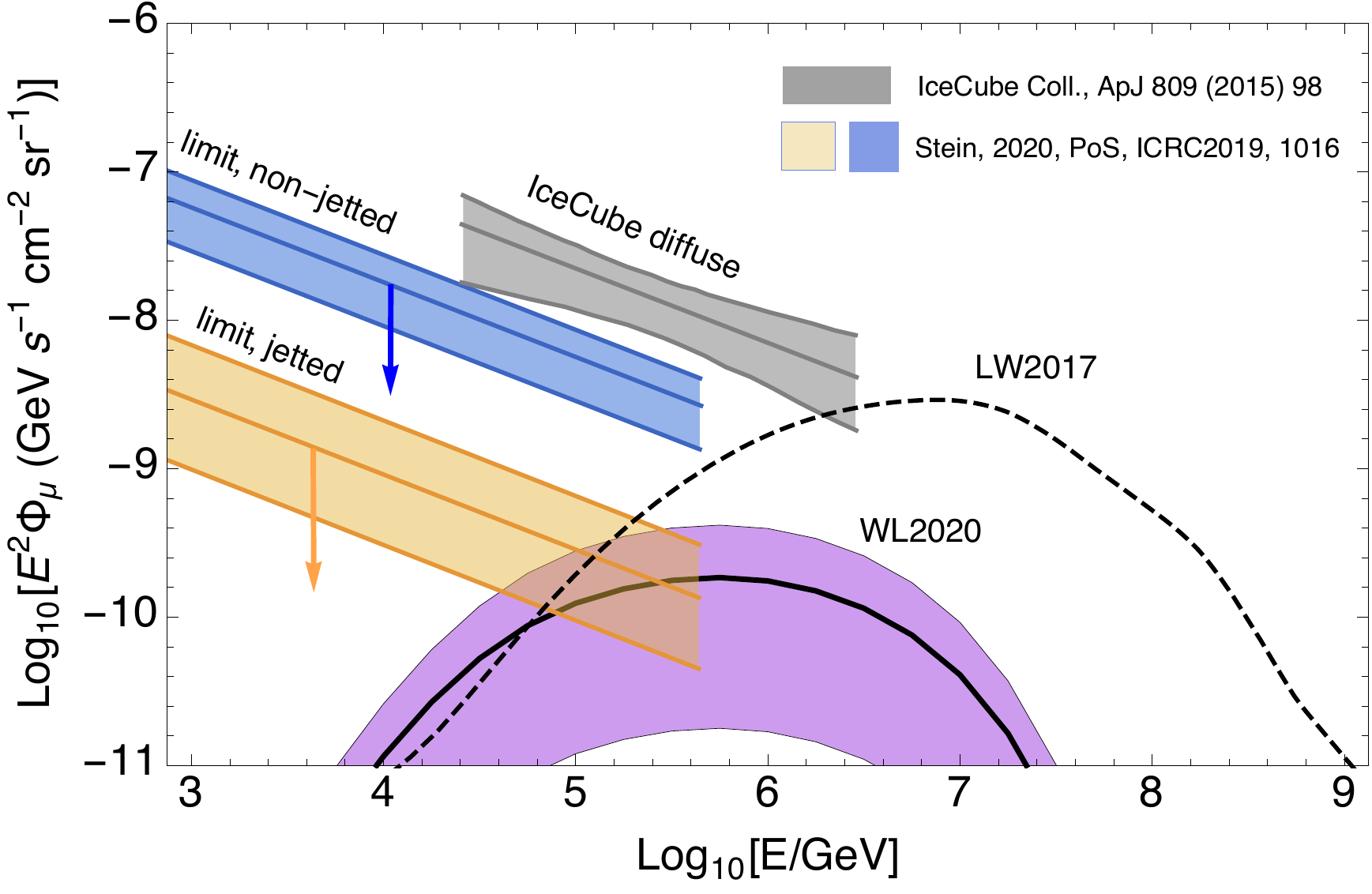}
    \caption{Diffuse flux expected from all TDEs for this model (purple region), the LW2017 model following Swift J1644+47~\cite{Lunardini:2016xwi} (dashed curve), in comparison to the observed diffuse astrophysical neutrino flux~\cite{Aartsen:2015knd} and two earlier TDE stacking limits~\cite{Stein:2019ivm}. The shaded area reflects the uncertainty on the minimum $M_{\mathrm{SMBH}}$ for which TDEs occur; the solid curve corresponds to $M_{\mathrm{SMBH}}\geq 10^5 M_\odot$; see \cite{Lunardini:2016xwi} for details.   }
    \label{fig:diffuse}
\end{figure}

It is interesting to re-consider what fraction of the diffuse neutrino flux detected at IceCube may come from TDEs in the light of the recent neutrino observation from AT2019dsg. Earlier models, based on the very luminuous jetted TDE Swift J1644+47 \cite{Burrows:2011dn}, predict a high diffuse neutrino flux \cite{Lunardini:2016xwi}, which is potentially in conflict with stacking and multiplet limits.  Instead, our current model, when extended to the whole TDE population, is consistent with these limits, and potentially leads to a few percent contribution to the diffuse neutrino flux.  This fraction is at the lower end of the range derived in Ref.~\cite{Bartos:2021tok} based on current observations. The uncertainty on the TDE diffuse flux is large, being dominated by the uncertain contribution of small mass SMBHs to the total TDE rate. A further uncertainty is due to the dependence of the neutrino production mechanism on $M_{\mathrm{SMBH}}$ (not included in Fig.~\ref{fig:diffuse}).  We also note that multiple contributions to the diffuse neutrino flux are expected for different reasons, such as the proper description of the energy spectrum~\cite{Palladino:2018evm}. 

Broadening the perspective to other messengers, studies have been done on TDEs as sources of UHECRs
~\cite{Biehl:2017hnb,Zhang:2017hom,Guepin:2017abw}, and a common origin of neutrinos and UHECRs has been proposed in~\cite{Biehl:2017hnb}. The main criticism to such scenario has been that the disruptions of stars with a heavier isotopic composition -- typically white dwarfs -- are needed to reproduce the observed UHECR composition.  Given the low rate of these disruptions, a potential tension with multiplet constraints was noted. The problem might be resolved by a new estimate of the rate of white dwarfs disruptions, which is a factor of 50 larger~\cite{Tanikawa:2021zfm}. This enhancement factor might restore the naturalness of the UHECR hypothesis, and may even reduce the required X-ray luminosity. 

As far as the advantages of the jetted model are concerned, we have highlighted energetics, timing (of the neutrino) and neutrino energy spectrum. In addition, we expect efficient particle acceleration in the jet, and our requirements for the acceleration efficiency are moderate. Although a relativistic jet has not been directly observed, there are some indications from optical polarimetry measurements~\cite{Lee_2020}, and even the radio signal may be interpreted as due to a relativistic jet in non-vanilla models~\cite{Cannizzaro:2020xzc}. In Ref.~\cite{Cannizzaro:2020xzc}, a late-term X-ray signal at about 115 days after discovery has been found as well, which does not fit the exponential decay trend described above; one may speculate that this could be a signature of the jetted radiation. It is also conceivable that there are macroscopic effects at work, such as jet precession, jet re-collimation, or twisting. These effects which decouple the neutrino and electromagnetic signatures, since these are expected to originate from different parts of the jet, see~\cite{Bustamante:2014oka}. 

\section{Summary and outlook}

We have discussed a theoretical intepretation of the neutrino observed from the TDE AT2019dsg. We have highlighted the delayed detection of the neutrino (with respect to the optical peak), its energy, and especially the overall scale of energy that is required for the neutrino fluence.  We have demonstrated that energy arguments require either a high SMBH mass and extremely efficient energy conversion into neutrinos, or  collimated emission -- such as from a relativistic jet, which we have followed here.  The observed neutrino energy and arrival time indicate a possible connection with the (observed) X-rays. Specifically,  the same effect that causes the decay of the X-rays (obscuration) may be lead to partial isotropization; the isotropized X-rays may then serve as external target for the jet. The parameters of the model are consistent with MHD simulations.
We find a neutrino light curve that is sustained over hundreds of days, consistently with the delayed (relative the optical peak) neutrino observation.  
The energy spectrum (multiplied by $E^2$) of the neutrinos is dominated by multi-pion processes; it is flat between about 100~TeV and 10~PeV. 
Although advantageous in many ways, the jetted interpretation remains only a possibility, since
no clear signature of a jet have been identified so far. Future observations will be needed to reach a conclusion in this respect.

After AT2019dsg, another neutrino event (IC200530A) from another TDE (AT2019fdr) has been observed, with an intriguingly similar time delay of a about 300 days with respect to the optical peak~\cite{SteinAtParis};   AT2019fdr was significantly more luminous than AT2019dsg. The similarities and differences between the two events will allow to identify trends 
in the neutrino production from TDEs, and therefore  to advance models significantly. We therefore expect new exciting results soon.

\bibliographystyle{JHEP}
\bibliography{references}

\end{document}